\documentclass[sigconf,authorversion]{acmart}

\renewcommand\footnotetextcopyrightpermission[1]{} % removes footnote with conference information in first column
\usepackage{xcolor}
\usepackage{multirow}   

\begin{document}
\fancyhead{}

%%
%% The "title" command has an optional parameter,
%% allowing the author to define a "short title" to be used in page headers.
\title{Mutltimodal AI Companion for Interactive Fairytale Co-creation}

%%
%% The "author" command and its associated commands are used to define
%% the authors and their affiliations.
%% Of note is the shared affiliation of the first two authors, and the
%% "authornote" and "authornotemark" commands
%% used to denote shared contribution to the research.

\author{Ruiyang Liu}
\affiliation{%
  \institution{ShanghaiTech University, SIMIT, UCAS}
  \country{China}
}

\author{Predrag K. Nikolić}
\affiliation{%
  \institution{ShanghaiTech University}
  \country{China}
}

%%
%% By default, the full list of authors will be used in the page
%% headers. Often, this list is too long, and will overlap
%% other information printed in the page headers. This command allows
%% the author to define a more concise list
%% of authors' names for this purpose.

%%
%% The abstract is a short summary of the work to be presented in the
%% article.
\begin{abstract}
\textit{AI fairy tale companions} plays an important role in early childhood education as an augmentation for parents' efforts to close the ``participation gap'' and boost kids' mental and language development. 
Existing systems are generally designed to provide vivid materials as unidirectional entertaining reading environments, e.g, visualizing inputting texts.
However, due to the limited vocabulary of kids, these systems failed to afford effective \textit{interaction} to motivate kids to write their own fairy tales.
In this work, we propose \textbf{AI.R Taletorium}, an illustrative, immersive, and inclusive multimodal AI companion, for interactive fairy tale co-creation that actively involves kids to create fairy tales with both the \textit{AI agent} and their \textit{normal peers}. 
AI.R Taletorium consists a neural story generator and a doodler-based fairy tale visualizer. We design a character-centric bidirectional connection mechanism between the story generator and visualizer equipped with Contrastive Language Image Pretraining (CLIP), thus enabling kids to participant in the story generation process by simple sketching. 
Extensive experiments and user studies show that our system was able to generate and visualize meaningful and vivid fairy tales with limited training data and complete the full interaction cycle under various inputs (text, doodler) through the bidirectional connection.
\end{abstract}


% \begin{teaserfigure}
%   \includegraphics[width=\textwidth]{figs/pipeline.png}
%   \caption{AI.R Taletorium prototyping. Following character-centric design we divide our system into three key functions: facial-attribute}
%   \label{fig:teaser}
% \end{teaserfigure}

%%
%% This command processes the author and affiliation and title
%% information and builds the first part of the formatted document.
\maketitle

\section{Introduction}
\begin{quote}
\textit{If you want your children to be intelligent, read them fairy tales. If you want them to be more intelligent, read them more fairy tales}.\\
\begin{flushright}
--- \textit{Albert Einstein}
\end{flushright}
\end{quote}

\begin{figure*}[t]
\begin{center}
   \includegraphics[width=1.0\linewidth]{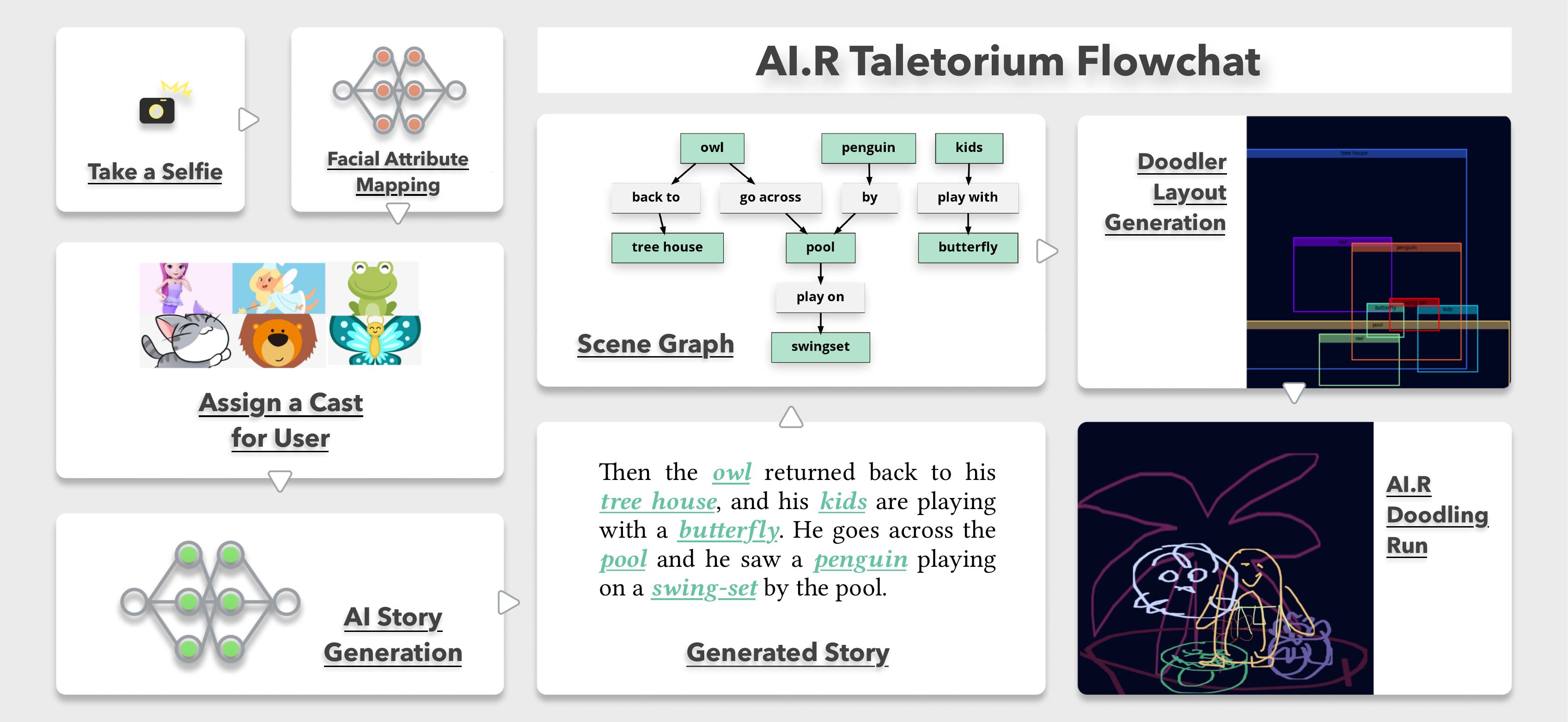}
\end{center}
\caption{System prototyping. The system is initialized with user facial features. We assign a fairy tale cast to each user participating (left). The AI fairy tale generator will generated an intial story based on casts (i.e characters). For visualization, we first parse the story into a scene graph based intermediate representation, then we adopt trained scene compositor to predict the layout for each cast in the story. Finally, we visualize the story by drawing doodlers according to the predicted layout.}
\label{fig:sys_proto}
\end{figure*}

Language and mental development are key aspects in early childhood education. Fairy tales, through the characters and virtues shown in the stories, provide moral models for kids to follow and help to develop their vocabulary by offering vivid illustrations. The majority of existing story visualization researches \cite{Gregor2015DRAWAR,Johnson,Johnson2018,Li2019ObjectDrivenTS,Liu2020,Rose2010} have relied on the combination of recurrent neural networks (RNN) and generative adversarial models (GANs). While achieving stunning results on constrained datasets like MS COCO~\cite{}, such end-to-end text-to-image mapping could fail on composite scenes with multiple objects. 
Recently, \cite{Herzig2019,Huang2019,Mirza2014} induce scene graph as an intermediate representation to support visualization of compound sentences. The Scene graph \cite{Johnson} describes things in a scene and their relationships, most commonly position relationships. Compared with raw text, scene graph conceptualizes scene contents as abstract semantic elements that are not bounded by object class or relationship types, boosting text-to-image studies in flexibility and representation power.
However, most existing visual language models are trained on massive real-life photos with captions. In fairy tale visualization, we still face the challenge of data shortage. On the one hand, fairy tale illustration conserves much artist labor; on the other hand, the illustration style usually varies from artist to artist, making it more challenging to learn. Based on this point, we combine a rule-based fairy tale parser with a neural sketch generator into an intelligent and interactive visualization system that specially focuses on fairy tale visualization.

In light of this, we propose a new AI fairy-tale co-creation system: AI.R Taletorium. 
xxx %todo 这里缺一块，针对上面的缺陷，我们针对性地提出xx，xx，xx分别解决上述的几个问题，取得什么效果，需要整理下
It enables common creative experiences between groups of children with different physical preferences and at other locations. We aimed at offering an entertaining and imaginative platform for children to communicate with each other, improving their psychological health, especially under extreme circumstances such as the COVID-19 pandemic. The system is composed by two novel modules, a character-centric story generator and a visualizer in the form of doodlers. We connect those two modules by characters and build a bidirectional link to allow kids to interactively participant in the fairy tale creation process by simple doodling without vocabulary requirements.

We summarize our key contribution as follows: 
\begin{itemize}
\item A novel multimodal fairytale co-creation interface based on interactive text-to-image transfer and vice-visa.

\item The system presents a novel fusion method between learn-based language model and rule-based graph update, allowing for more flexibility on both story generation and visualization with limited data.

\item Proposing a CLIP transformed story graph as an intermediate representation to transcends the barrier between digital storytelling content and insufficient illustration data (automatic AI fairy tales storytelling case).
\end{itemize}

% todo：给一下章节的引用
In this paper, we will first give a brief description of the AI.R. Taletorium system. Then we explain our character-centric approach for fairy tale generation and visualization. 
Finally, we present the results from the performed experiments and demonstrate various applications to prove the generation flexibility of our proposed system.

We leverage AI’s potentials in both knowledge understanding and creation process to design a co-creative environment for imaginative Human-AI interactive storytelling. The fairy tale storyline and content illustrations elicit AI interpretation of the user’s facial characteristics and visual content. They are adding during the process of AI agent fairy tales generation. AI.R Taletorium is a multi-user platform, which runs online and follows the basic principle of \textit{PWA (progressive web application)} design. It works in real-time on multiple devices, which enables users to participate at any time from anywhere.

In this paper, we focus on AI.R Taletorium intelligent and artistic fairy tale visualizer that interacts with users in the real-time proposal. While the story provides an effective and concise way to share experiences, visual content acts as a more comprehensive and universal communication tool that transcends barriers between user groups with different cultural backgrounds and psychological and physical preferences. The learning for meaningful and coherent story visualization thus becoming a popular yet challenging task among AI studies.

Story visualization as a downstream task of text to image generation has met with great success in the deep learning era. Most studies \cite{Gregor2015DRAWAR,Johnson,Johnson2018,Li2019ObjectDrivenTS,Liu2020,Rose2010} have relied on the combination of recurrent neural networks (RNN) and generative adversarial models (GANs). While achieving stunning results on constrained datasets like MS COCO, such end-to-end text to image mapping could fail on composite scenes with multiple objects. Recently, \cite{Herzig2019,Huang2019,Mirza2014} induce scene graph as an intermediate representation to support visualization of compound sentences. The Scene graph \cite{Johnson} describes things in a scene and their relationships, most commonly position relationships. Compared with raw text, scene graph conceptualizes scene contents as abstract semantic elements that are not bounded by object class or relationship types, boosting text-to-image studies in flexibility and representation power.

However, most existing visual-language models are trained on massive real-life photos with captions. In fairy tale visualization, we still face the challenge of data shortage. On the one hand, fairy tale illustration conserves much artist labor; on the other hand, the illustration style usually varies from artist to artist, making it more challenging to learn. Based on this point, we combine a rule-based fairy tale parser with a neural sketch generator into an intelligent and interactive visualization system that specially focuses on fairy tale visualization.

We summarize our key contribution as follows: 
\begin{itemize}
\item Novel approach to multimodal interface design based on dynamic AI text-to-image visualization.
\item Proposing intelligent visualization interface that enables content co-creation between users and AI.
\item Proposing a story graph as an intermediate representation to transcends the barrier between digital storytelling content and insufficient illustration data (automatic AI fairy tales storytelling case).
\end{itemize}

The system presents a novel fusion method between a learn-based language model and rule-based graph update, allowing for more flexibility in defining visualization flexibility.

First, we will give a brief description of the AI.R. Taletorium consisting of parts. We will then explain our approach to fairy tales visualization, the patterns that define AI as an evolved visualization interface, and present the results from the performed experiments. Lastly, we will conclude and specify future directions of the project.
\section{Related Work}

In the field of text to visual content(or, more specifically, image) synthesis \cite{Gregor2015DRAWAR,Herzig2019,Johnson2018,Koh2020TexttoImageGG,Li2019ObjectDrivenTS,Mansimov2016GeneratingIF,Ramesh2021ZeroShotTG,Reed2016GenerativeAT,Schuster2015GeneratingSP,Yan2016,Zhang2017StackGANTT,Zhang2019StackGANRI}, early studies \cite{Zhu} usually ground text-to-image generation as a search problem. Given a text description, they search for words, representative images, and placement from an existing database that most match the description. They then adopt a graphics renderer to combine the search results into the same canvas. Despite the complex AI models (vision, language, and graphics) integrated into the system, such systems are usually incapable of generating new images. 

With the rise of generative models\cite{Goodfellow2020,Mirza2014}, researchers also explore combining RNN and GANs for text-to-image generation by learning directly from massive text-image pairs \cite{Gregor2015DRAWAR,Li2019ObjectDrivenTS,Koh2020TexttoImageGG,Mansimov2016GeneratingIF,Ramesh2021ZeroShotTG,Reed2016GenerativeAT,Zhang2017StackGANTT,Zhang2019StackGANRI}. Such modelling enables the generation of new visual content from concise text descriptions. However, constrained by training data formation, they are generally incapable of modeling complex scenes with multiple objects.

Another stream of work adopts intermediate representations to deal with complex scenes. Johnson et al.\cite{Johnson2018} firstly introduce scene graph for image generation and outperforms previous methods in generating an intricate image with multiple objects. They develop a graph convolution network to process the scene graph, which acts as a principal component for following scene graph-based image generation methods \cite{Herzig2019,Schuster2015GeneratingSP}.

Similarly, we use scene graph as the medium between fairy tale and visual contents to support sketched scene generation from fairy tale fragments with unusual objects and complex relations. Along with the generation of fairy tale fragments, we par each fragment in real-time into a scene graph containing characters and their relationships covered by the fragment. We then propose a conditional neural scene composer learned from human captioned natural images to compose sketched scenes from the parsed scene graph. As the story goes fragment by fragment, we update the scene graph according to fragment co-references \cite{Sukthanker2020AnaphoraAC}.

\subsection{Learning for sketching}
Doodling, or free-hand sketch, is a universal communication and art modality which combines convenience with expressiveness. As a high-level abstraction of real-world contexts, doodling endorses advantages for neural processing with its simplicity while suffers from the high sparsity and wide style diversity. Based on neural representations for sketch structure\cite{Ha2018ANR,Ribeiro2020SketchformerTR}, a great amount of work has been done in doodle creation, recognition, retrieval, partial analysis and, abstraction, etc. We refer the audience to\cite{Xu2020DeepLF} for a comprehensive review of learning for sketching. 

With artificial intelligence becomes increasingly part in people’s everyday lives, making researches on human-AI co-creation rather essential. Recently, \cite{Ostrowski2020DesignRI} propose the positional idea to grounds the HRI design research on three touch-points: the roboticists as designers, the design features of the systems, and the users as co-designers. \cite{Parikh2020ExploringCC} explore four scenarios for collaborative human-human sketch co-creation, and shows that collaborative drawing with a third-party voting strategy leads to most creative sketches. Despite these pioneering works, the research in human-robot co-creation remains extremely limited. With AI.R Taletorium, we proposed an efficient human-robot co-creation scenario that utilize AI as a powerful interface.

The visualization system of AI.R Taletorium combines doodle creation with comprehension. It not only enables doodling from the story but also turns the user sketch interactively into the story.

\subsection{Interactive Storytelling (IS)}
Interactive storytelling plays an important role for for Early Childhood Education. AI learning companions, as either tutor or tutee role, provides an intelligent and responsive interface to support children’s learning in a variety of contexts including language development, storytelling and scientific learning \cite{Kanda2004InteractiveRA,Park2019AMA,Park2017TellingST,Gordon2016AffectivePO,Chen2020ImpactOI}. Compared with kindergarten teachers, personalized AI tutors/tutees could asses and automatically adapt to kid's knowledge level and physical needs. Previous studies\cite{Belpaeme2018SocialRF,Park2019AMA,Maeda2019CanAS} have emphasized that the personalized instruction from robots could clearly augmenting the efforts of parents and teachers to help kids acquiring both academic knowledge and positive attitudes. 

Storytelling, as an interactive process that facilitates imagination, creative thinking, language abilities, and cooperative learning, bringing a broad range of positive outcomes for primary education \cite{Garzotto2014InteractiveSF} summarizes methodologies and enabling technologies used in building interactive storytelling system for kids. The In-Visible island project \cite{Talib2020InVisibleII} further pinpoints digital inclusion in IS systems. They designed an inclusive system to join visually impaired children and sighted kids together into a unified storytelling process, which on the one hand, accelerated the social learning curve of visually impaired children, on the other hand, educating sighted children to have more empathy towards their peers with physical limitations.

We design AI.R Taletorium to be character-driven. In AI.R Taletorium, the story is generated based on characters\cite{Liu2020,Surikuchi2019CharacterCentricS}, the scene graph is updated based on character coreferences, and the interaction is based on users adding/removing doodler characters from the drawing canvas.

\section{AI.R TALETORIUM SYSTEM}

\begin{figure*}[t]
\begin{center}
   \includegraphics[width=1.0\linewidth]{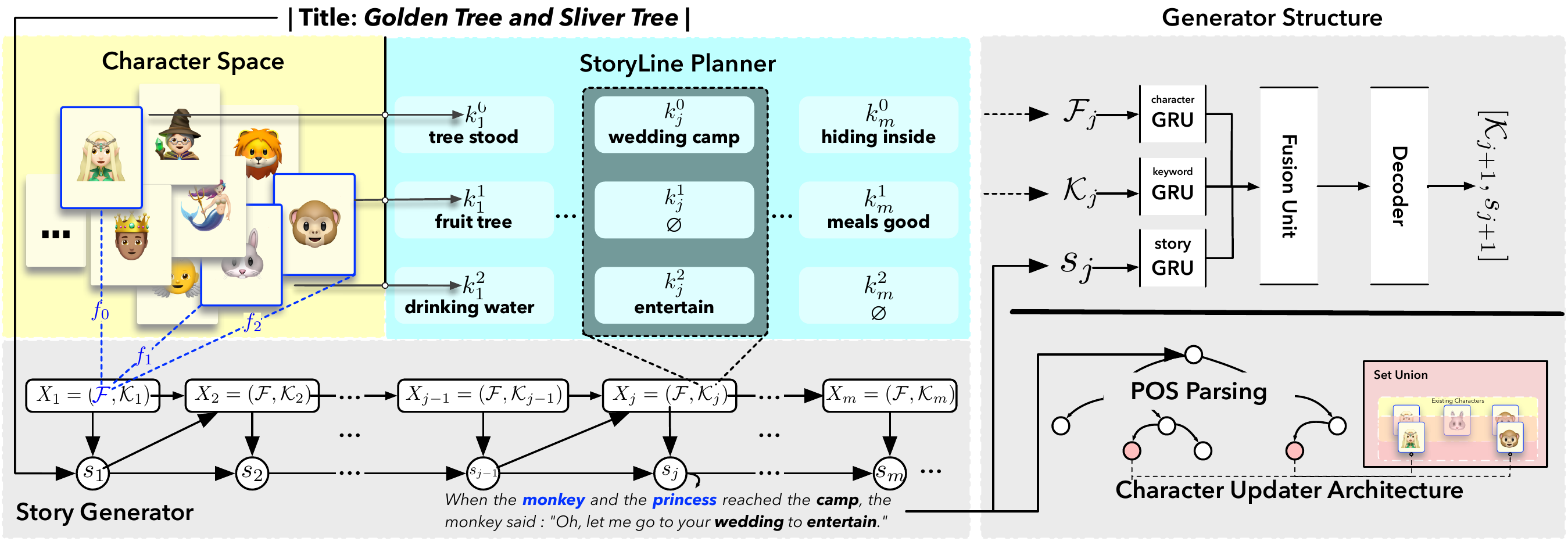}
\end{center}
\caption{Character-centric story generation. We combine plan-and-write\cite{Yao} story generation framework with character features to compose each story fragment. We apply a dynamic generation strategy and plan for each fragment on the fly based on the previous story fragment and current character status. Such a character-centric generation scheme thus supports user co-creation with AI in storytelling by modifying characters.}
\label{fig:pnw_net}
\end{figure*}

AI.R Taletorium is a character-centric multi-modal AI fairy tale telling system aims at connecting users with different background and physical needs into a unified fairy-tale co-creation process with AI. The system consists of two intelligent components:
\begin{itemize}
    \item Character-centric AI fairy tale generator
    \item Intelligent doodling-based fairy tale visualizer. 
\end{itemize}
The system interacts with the users in the following ways:
\begin{itemize}
    \item Collecting users’ facial characteristics and selecting characters for the storyline based on predefined matching criteria.
    \item Interpreting users’ doodlers, they are added to the AI generated fairy tale illustration and creatively implementing it in the storyline.
\end{itemize}

With AI.R Taletorium, we aim to intensify User-AI interaction in the joint act of creation by using users’ personal properties (face features and doodlers) and challenge artificial intelligence creativity to the highest level of imagination.

\subsection{AI.R Tentorium Character-Centric Fairy Tale Design Approach}
We formulate the system input\/output as follows:

\textbf{Input:} A character set $\mathcal{C}=\{c_0,c_1,..., c_{n-1}\}$, a title $\mathcal{T}=\{t_0,...,$ $t_{k-1}\}$ that defines the main topic of the fairy tale and a storyline size $m$.

\textbf{Intermediate output:} We map characters onto the character embedding space as character feature vector $\mathcal{F}=\{f_0,f_1,...,f_n\}$ as in \cite{Liu2020}. At each generation step $j\in[1,...,m]$, we generate storyline keywords\footnote{We allow $k_j^i$ to be empty, which means the character is not participating the current step.} $\mathcal{K}_j=\{k_j^0,k_j^1,...,k_j^n\}$ for each character according to previous story fragment to define character actions at current step.

\textbf{Output:} A story $S=\{s_0,s_1,...,s_{m-1}\}$ composed with $m$ story fragments, where $s_j=Merge(l_j^0,l_j^1,...,l_j^n)$ is merged from $n\leq\|\mathcal{C}\|$ sentences\footnote{We define $l_j^i=\emptyset$ when $k_j^i=\emptyset$, which means some characters might not be involved from the current generation step, and current fragment size n could be less than number of characters $\|\mathcal{C}\|$.} generated from each character’s storyline keyword at current step.

We follow a recurrent generation framework\cite{Yao} to generate the whole story (Fig.~\ref{fig:pnw_net}). Each fragment is generated based on previous fragment and current status $X_j=(F,K_j)$. We initialize the first fragment as the given title:

\begin{equation}
s_{j+1}=(s_j,X_{j+1}),\quad s_0=\mathcal{T}={t_0,t_1,...,t_k}      
\end{equation}

As shown in Fig.\ref{fig:sys_proto}, we initialize the characters by mapping user facial attributes onto 24 predefined fairy tale characters according to \textit{Psychological $\&$ Cultural Stereotypical} rules.

\begin{figure}[t]
\begin{center}
   \includegraphics[width=1.0\linewidth]{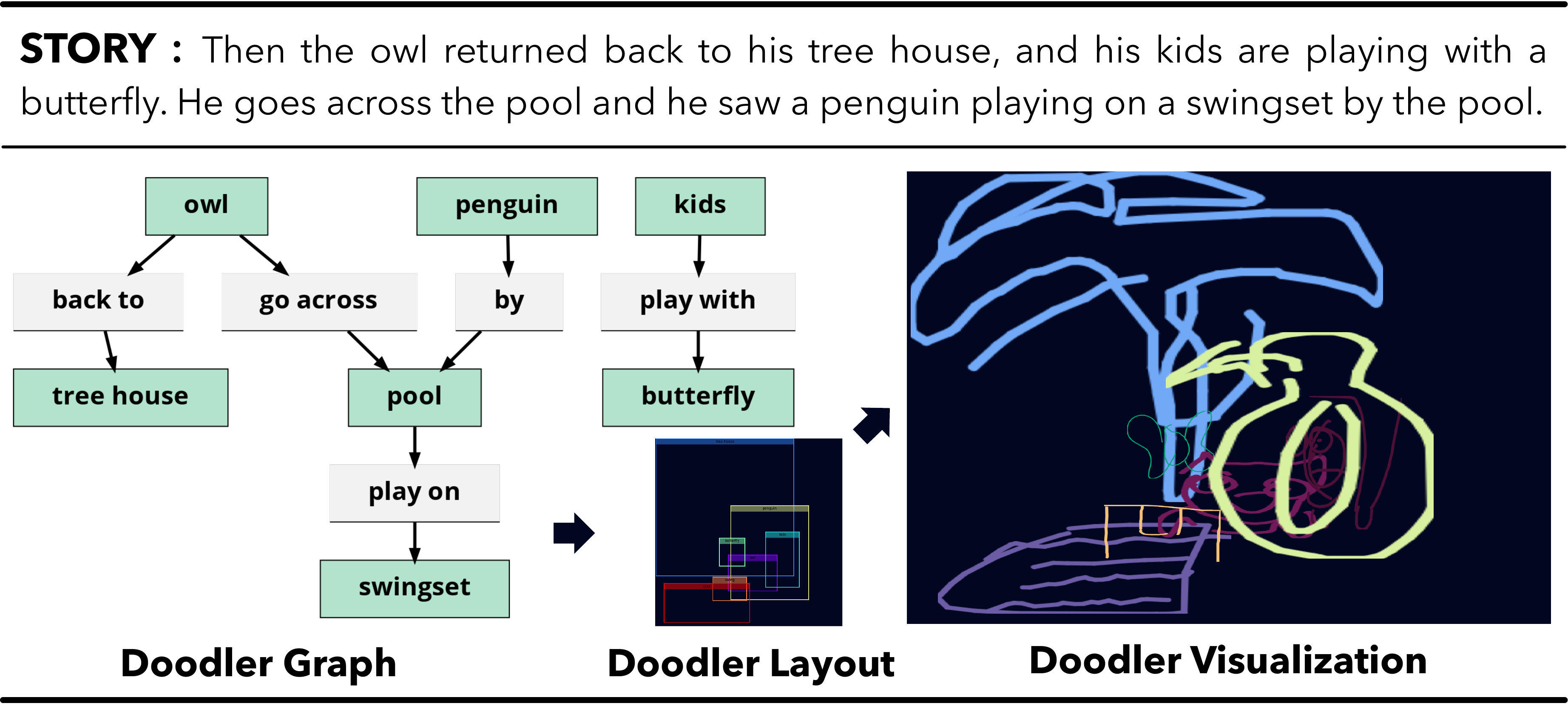}
\end{center}
\caption{Fairy tale fragment to doodler visualization. Given a fairy tale fragment (top), we first adopt a rule-based parser to turn it into a doodler graph (left). Then we predict the doodler layout from the doodler graph (middle). Finally, we compose the doodler visualization according to the layout (right).}
\label{fig:dg_pipeline}
\end{figure}

\section{Intelligent Fairy Tale Visualization}

Visualization plays an important role for people, especially kids, to understand and enjoy fairy tales. Following character-centric design, in AI.R Taletorium, we adopt AI to automatically generate an interface based on fairy tale characters that allow for direct user interaction.

Given a story fragment, AI.R Taletorium automatically builds an interactive canvas and draws each character based on their spatial relationship as doodlers on the canvas. We address the canvas as an intelligent interface, within which each doodler builds a button that supports the user to interact with the story generator by adding or removing doodlers.

Most state-of-the-art text-to-image visualization methods are based on data-intensive transformer architecture, which generally suffers from insufficient training data, not to mention the domain gap of language styles and object categories between fairy tale stories and formal language descriptions. For example, while ``A horse under a tree'' could be a formal description for Visual Genome dataset\cite{Krishna2017}, in fairy tale stories it becomes \textit{``The ferocious unicorn trapped in the tree''}. 

To tackle this problem, we introduce a doodler graph as an intermediate representation and adopt a two-stage generation framework to visualize fairy tale stories to interactive doodling interfaces. As shown in Fig. \ref{fig:dg_pipeline}, we first parse the fairy tale fragment description into a doodler graph, then we adopt a GNN scene composer to compose doodler visualization from the parsed doodler graph. Similar to \cite{Huang2019}, the scene composer will first predict character layouts, then we adopt the SketchRNN \cite{Ha2018ANR} decoder to draw individual sketches according to the layout. By introducing the doodler graph as an intermediate representation, our visualization system is able to deal with more complex and abnormal scene structures such as fairy tale scenes. In the following, we will first formulate the doodler graph representation, introduce the scene composer architecture and demonstrate the doodler graph updating policy during the story generation process.

\begin{figure}[t]
\begin{center}
   \includegraphics[width=1.0\linewidth]{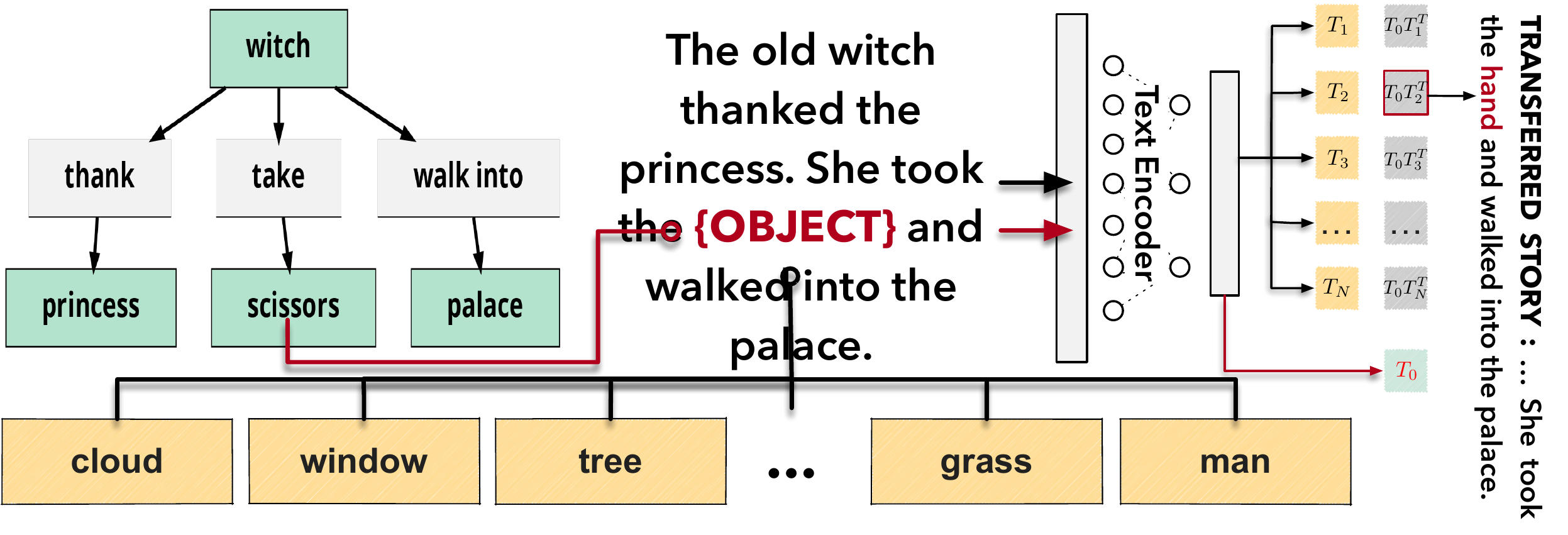}
\end{center}
\caption{Mapping doodler graph to VG scene graph. For characters and relationships presented in the doodler graph, we adopt a pre-trained text encoder\cite{Radford2021LearningTV} and search for the most similar VG\cite{Johnson} object based on the current story context.}
\label{fig:dg2sg}
\end{figure}

\begin{figure*}[!htb]
\begin{center}
   \includegraphics[width=0.9\linewidth]{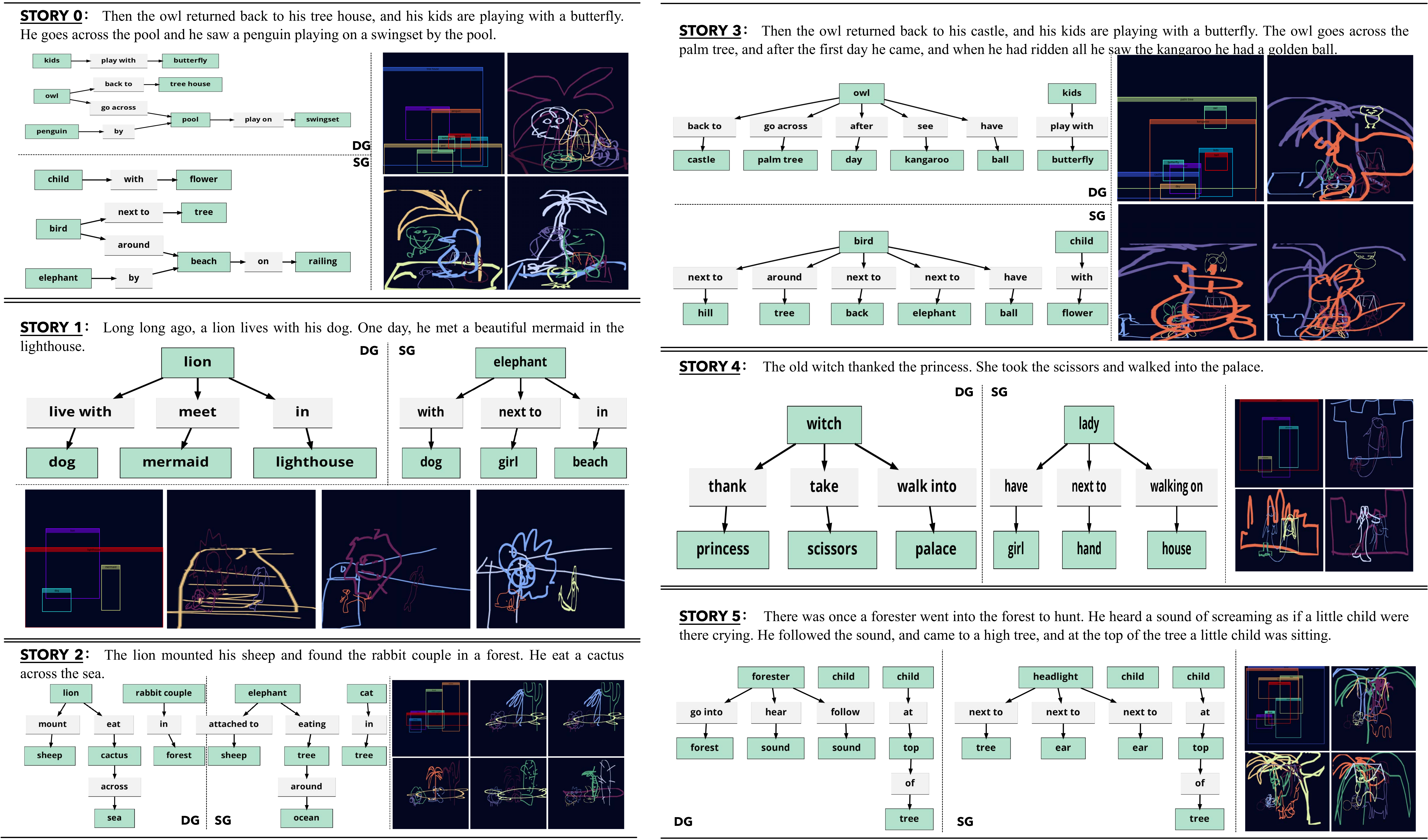}
\end{center}
\caption{Doodler scene composition results. We demonstrate composed doodler scene examples for 6 stories. For each story, we show the parsed doodler graph (DG), the transformed casual scene graph (SG) based on the mapping strategy described in Sec. \ref{sec:scene_composer}(Fig.\ref{fig:dg2sg}) together with 4 composed scene examples. The result shows that with the doodler graph representation, we could handle complex scenes with multiple objects and relationships. Additionally, comparing STORY 3 with STORY 0, we further shows that our doodler graph representation could automatically remove redundant information and non-sense AI generated sentences like \textit{\'when he has ridden all\'}. Meanwhile, as the mapping strategy automatically turn fairy tale objects to casual scene graph objects, we could also deal with rare scene descriptions including \textit{\'witch\'} and \textit{\'forester\'}.}
\label{fig:res_gen}
\end{figure*}

\subsection{Doodler Graph Representation}
Same as scene graph, doodler graph is a graph data structure that describes the contents of a scene\cite{Johnson,Johnson2018,Johnson2018ImageGF}. We formulate the doodler graph as $G=(V,E)$, where vertexes $V={v_1,...,v_n }$ refers to the set of characters, and edges $E$ represents character relationships. Each doodler character $v_i=(o_i,c_i,w_i)$ is identified by its category $o_i$, color $c_i$ and stroke weight $w_i$. 

Given a story fragment, we adopt a rule-based semantic parser to generate the doodler graph. Similar to\cite{Johnson2018,Wu,Schuster}, the parser will first parse the fragment into a syntactic dependency tree. Then a set of tree transformations is applied to the dependency tree to acquire characters (i.e., nouns), adjectives that describe character attributes, and verbs that identify character relationships. In practice, we assign doodling color and stroke weight to each character either based on their attributes (e.g., a red flower) or randomly choose one from a color palette.

\subsection{Doodler Scene Composer}
\label{sec:scene_composer}
We train the doodler scene composer directly on the Visual Genome dataset (VG) \cite{Krishna2017}. Considering the gap of characters and relationships between fairytale scenes and real-world scenes, we adopt a pre-trained text-encoder \cite{Radford2021LearningTV} to map the doodler graph into a scene graph according to the current story fragment. Specifically, we treat the matching problem as a zero-shot prediction problem. We replace each doodler entity in the story fragment with 81 candidate VG entities. We then encode candidate story fragments into text features ${T_1,T_2, ..., T_N}, N=81$ and calculate their scale cosine similarity with original story fragment feature $T_0$, we choose the most similar entity to replace the doodler entity (Fig. \ref{fig:dg2sg}). We adopt the same way to transfer doodler relations into casual relations. With transferred doodler entity and relation, we build a casual scene graph that suits the VG dataset.

After transferring the doodler graph to scene graph, we adopt a graph-neural-network-based layout predictor \cite{Johnson2018} trained on VG dataset to predict layout for doodlers. Given the layout bounding box, and layers for each object, we adopt a modified Sketch-RNN decoder \cite{Huang2019} to draw each doodler entity onto the canvas. 

% \subsubsection{Perspective scene composer}

\subsection{Doodler graph update policy}
As the story going fragment by fragment, we update the doodler graph based on two policies with respect to stories and characters:
\begin{itemize}
    \item The \textit{\textbf{story-based policy}} is based on neural co-relation analysis and refreshes the doodler graph when a new story fragment comes. 
    \item The \textit{\textbf{character-based policy}} adopts a learned graph modifier and performs nodes insertion and deletion regarding adding/removing characters from the scene.
\end{itemize}

\subsubsection{Story-based policy}
As the story goes fragment by fragment, we update the doodler graph with co-reference analysis followed by entity linking. The co-reference system keeps track of pronouns in the system and matches the pronouns to nouns that existed in the previous context. We thus use the co-reference to link doodler characters in the newly generated fragment to the existing fragment. We analyze co-reference between adjacent story fragments. After the co-reference is resolved, we update pronouns to their referenced nouns. With the updated doodler graph, we predict layouts for new doodler characters and add each new character individually onto the canvas.

\subsubsection{Character-based policy}
The character-based update policy is similar to the story-based policy and is used to support user interaction. As a visualization medium, the doodler graph supports both transformings into/from story fragments and doodling. AI.R Taletorium is integrated with a doodler recognition system and allows users to add characters along with the generation. We will automatically parse user sketch into story characters and update the scene graph based on doodler spatial relationships. We'll generate the next story fragment based on the updated character set.
\begin{figure*}[!htb]
\begin{center}
  \includegraphics[width=1.0\linewidth]{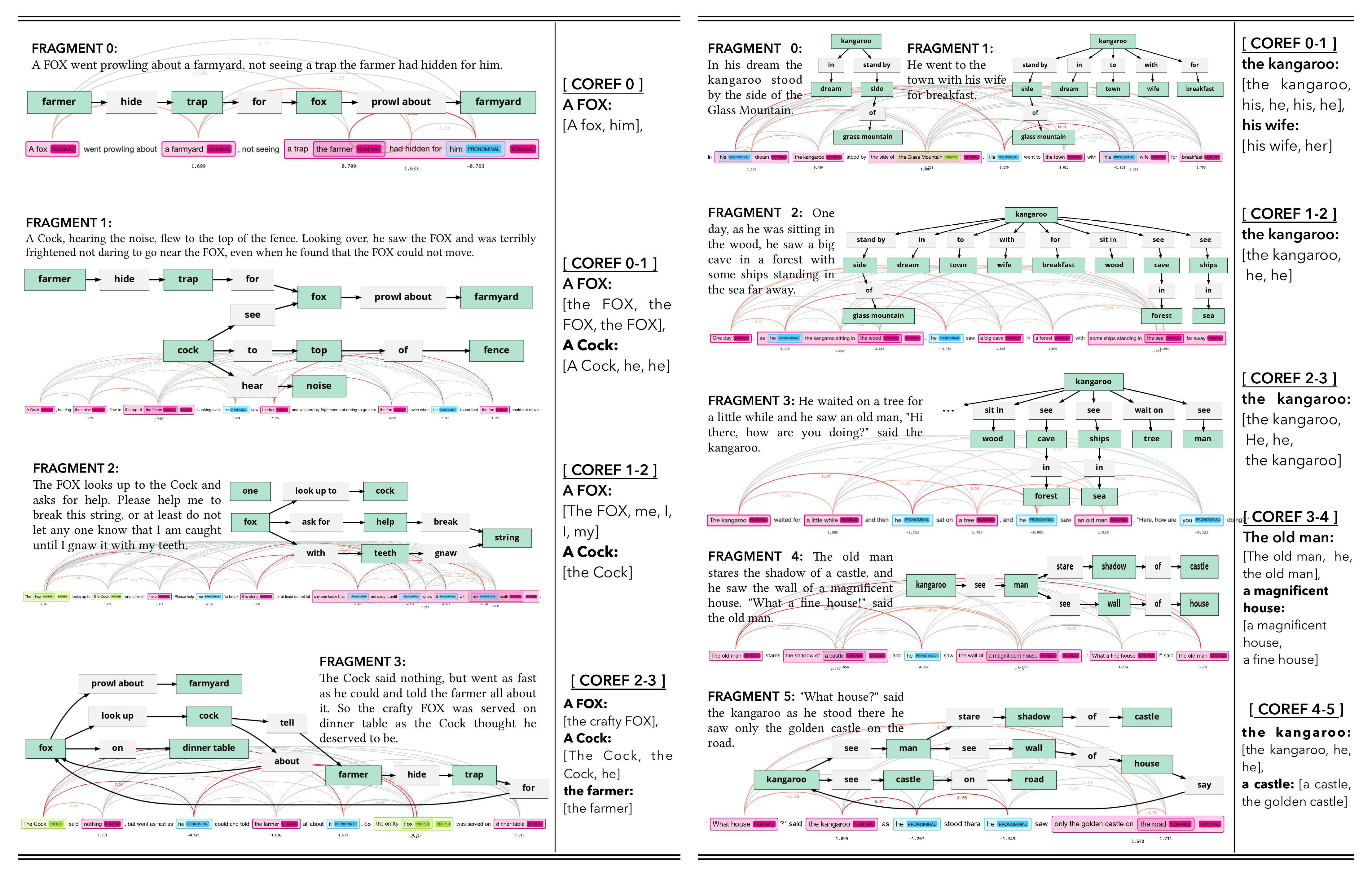}
\end{center}
\caption{Story-based doodler graph update of Aesop's fable story (right) and AI-generated story. Instead of generating a new doodler graph at each story fragment, we update the graph along with the generation to keep the visualization consistency (i.e., character consistency) within the same story.}
\label{fig:res_update}
\end{figure*}

\section{EXPERIMENTS}
\label{sec:expr}

In this section, we will first introduce the datasets we use and our implementation details in Sec \ref{sec:dataset-implementation}. We then conduct the qualitative evaluation with story visualization and demonstrate user interactions under three scenarios in Sec \ref{sec:interface-creation}.

\subsection{Dataset and Implementation} \label{sec:dataset-implementation}

Our system consists of three trainable modules: the AI fairy tale generator, the doodler scene compositor , and the sketch drawer. We use the following datasets to train each module:

\begin{enumerate}
    \item \textbf{Fairytale Dataset (FT890):} We collect 890 fairy tale stories to train the AI fairy tale generator. We applied a modified RAKE \cite{Rose2010} to generate storyline keywords for each story. Firstly, we divide each story into an appropriate number of fragments (5 during the experiment). We then extract a set of keywords from a story and get their fragments id and score (indicates its importance calculated by RAKE). We pick the top 5 keywords as the storyline for each fragment. For each story, we select the top 4 characters to build the character feature space according to storyline keywords. We train the story generator with two 3090Ti for 20 hours and test with the 500 epoch checkpoint during inference. \\
    
    \item We train the doodler scene composer on \textbf{Visual Genome Dataset} \cite{Krishna2017} which contains 81 object categories. We process each ground truth scene graph with a graph convolution network (GCN) and train the GCN to predict the bounding box conditioned on the scene graph. During inference, we map each fairy tale doodler to a scene graph object based on their similarity described in Fig. \ref{fig:dg2sg}. \\
    
    \item We train our doodler drawer on \textbf{QucikDraw Dataset}\cite{Ha2018ANR} which contains 345 object categories. For fairy tale characters beyond the dataset, we assign the doodler that is semantically similar to the character as its representative doodler. We adopt a trained  SketchRNN\cite{Ha2018ANR} on the dataset to draw individual doodlers. We utilize TFJS integration for front-end sketching.
\end{enumerate}

Our system is able to run in real-time during inference. With generated fairy tale, we adopt a rule-based doodler graph parser to generate the doodler graph and then map it to a casual scene graph for composition analysis. We build a web server to streaming the composition information to the front-end doodler drawer for visualization. We combine the doodler drawer with a sketch recognizer so that the user could interact with the AI.R visualizer by adding doodlers onto the canvas. We transfer the user drawing into the doodler graph with a character-based updating policy to guide the story generation.

\begin{figure}[!thb]
\begin{center}
   \includegraphics[width=0.9\linewidth]{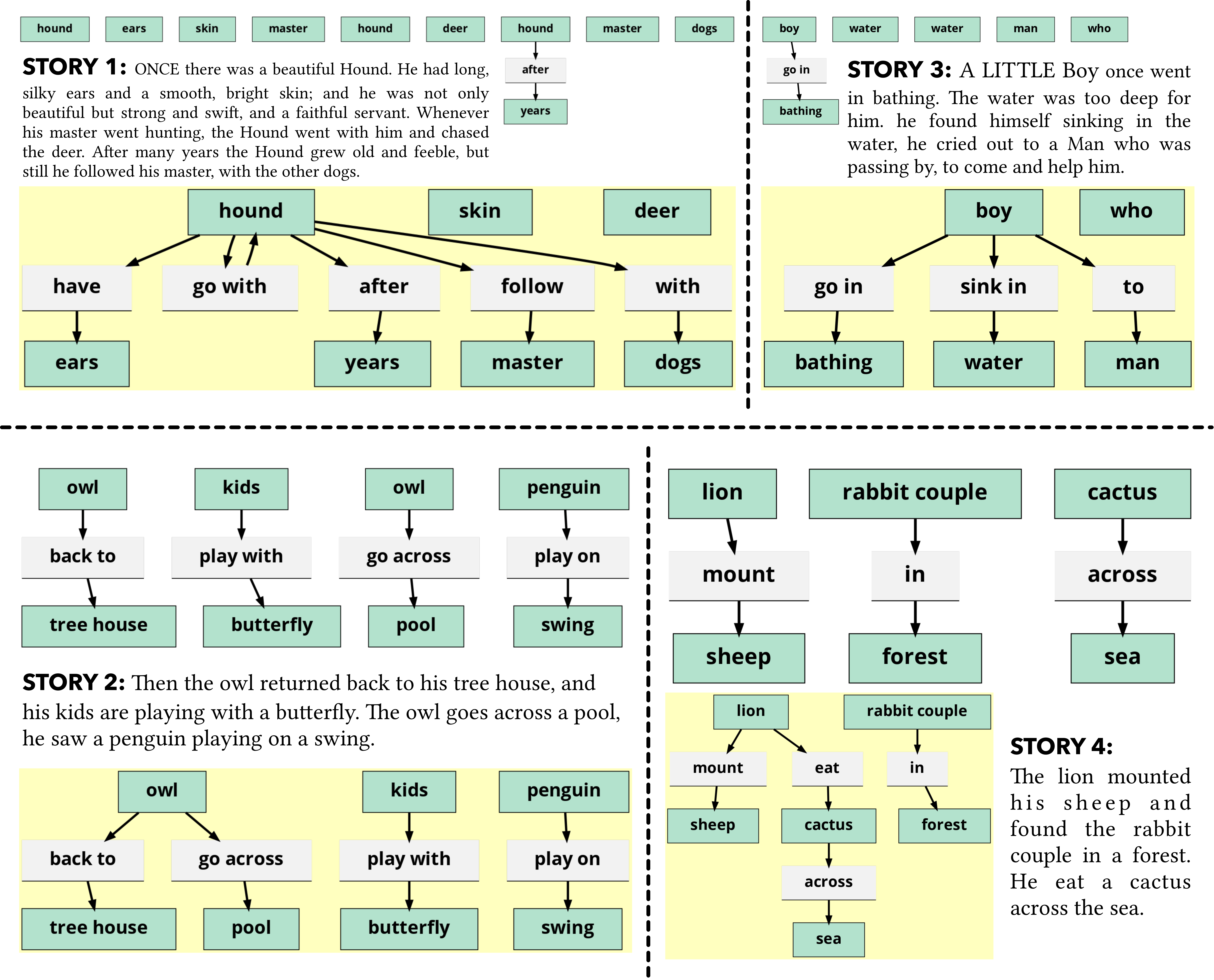}
\end{center}
\caption{Graph parsing w/o coreference resolution. We demonstrate the graph parsing result on two types of text inputs: Aesop store (top) and AI-generated fairy tale (bottom). We label the coreferenced graph with yellow.}
\label{fig:coref_comp}
\end{figure}

\subsection{Intelligent Interface Creation} 
\label{sec:interface-creation}

In the following, we first demonstrate the doodler scene composition results with parsed doodler graph, transformed casul scene graph and generated composition examples. We then analysis the efficiency of doodler graph updating policy among story fragments and perform ablation study on graph parsing w/o coreference resolution and scene composition with simple text guidance.

\subsubsection{Doodle Scene Composition}

We demonstrate the visualization result for fairy tale fragments in Fig. \ref{fig:res_gen}. Compared with end-to-end text guided sketch composition methods\cite{Huang2019}, with the doodler graph as intermediate representation, we are able to catch complex relations between multiple characters and scene objects, and identify correctly the layout for each object in the final composition with layered effects, i.e., the background scene could be correctly placed under character layers. For example, in the first story (story 0), there are 2 sentences with 7 scene characters/objects (green nodes), including an owl with his two kids playing with a butterfly and a penguin on swing-set by the pool, the doodler graph could correctly identify the relationships between characters and with the scene compositor we could align each character in a reasonable way (right). 

The doodler graph representation also tolerates a certain degree of grammar mistake and non-sense descriptions, which is a common case for AI generated stories. For example, story 3 has relatively the same objects with story 0, except the additional description like \textit{\'when he had ridden all\'}, the doodler graph automatically gets rid of the redundant information and the finally composited scene is similar to the original story (story 0 in this case). 

 AI Creativity plays an important rule in most generation tasks, especially in our case, when AI is acting as a co-creator for kids. During experiment, we find that the doodler graph representation also empowers the AI sketcher with certain degree of rational imagination by simple object/relation mapping to real world datasets (e.g. the VG dataset\cite{Krishna2017}). For example in story 5 (Fig. \ref{fig:res_gen}), while the \textit{\'forester\'} seemed to be unfamiliar for most AI interpreters, we automatically map it to a \textit{\'headlight\'} that is in the training dataset and favors the scene description.

\subsubsection{Doodler Graph Update}
As the story is generated fragment by fragment, making it rather important to keep visualization consistency between story fragments. Instead of generating new doodler graph for each story fragment, we automatically update maintain a single doodler graph for each story and update the graph based on neural coreference resolution across fragments. We assign for each character in the story a unique id and keep tracking of the character action in newly generated story fragments. In Fig. ~\ref{fig:res_update}, we analysis the automatic doodler graph updating policy with both human-written Aesop Fable stories (left) and AI-generated fairy tales (right). 

Aesop Fables are short stories composed by strongly connected fragments. In the example story, the \textit{\'FOX\'} is identified as the main character who will lead the story in multiple story fragments. With coreference resolution (bottom graph with red nodes in each example in Fig. \ref{fig:res_update}), we're able to track the same \textit{\'FOX\'} within multiple story fragments and update it's action with multiple supporting roles (e.g. the \textit{\'cock\'}).

Another problematic situation we dealt with the doodler graph updating policy is that the same character may perform in distant fragments. For example, though the \textit{\'farmer\'} only comes in the first and last story fragments, we managed to keep him alive through the whole story, and identify he's the same \textit{\'farmer\'} that the \textit{\'cock\'} speaks to in the last story fragment.

The rational connection among AI generated story fragments are comparably weak than human-written stories, Fig. ~\ref{fig:res_update} (right). Though according to the AI generated story, the \textit{\'kangaroo\'} performs multiple disconnected tasks amoung all scenes, with the updating policy, we could correctly track him as the main character of the story, and update supporting roles to the scene based on their relationships with the \textit{\'kangaroo\'}.

\subsubsection{Ablation Study}
We compare the generated doodler graph w/o coreference based updating in Fig. \ref{fig:coref_comp}. The coreference resolution helps in two ways. 

First of all, it helps to identify cross-fragment character relations, and keep tracking of the character action for long stories. For example, in the first story, the original scene graph parser failed to identify the  \textit{\'hound\'}'s activities and relations with supporting roles across the story while with the coreference resolution, we could correctly catch the \textit{\'hound\'}'s occurrence among the whole story to build the doodler graph. The same situation also suits for the \textit{\'boy\'} in story 2 and the \textit{\'lion\'} in the story 4.

Secondly, the coreference resolution helps to identify the same character occurred in distant story fragments. Thus, avoiding repeated visualization for the same character, and keep the style consistency for each character during the whole story and meet the real-time needs for interactive story telling. For example, while the original scene graph parser generates two \textit{\'owl\'} characters, we correctly identity them as the same \textit{\'owl\'} with coreference resolution, which also helps to build the relation that the same \textit{\'owl\'} going across the pool and meet the penguin. We show examples of composed scene for this story in Fig. \ref{fig:res_gen} (story 0).

\begin{figure}[t!]
\begin{center}
   \includegraphics[width=0.9\linewidth]{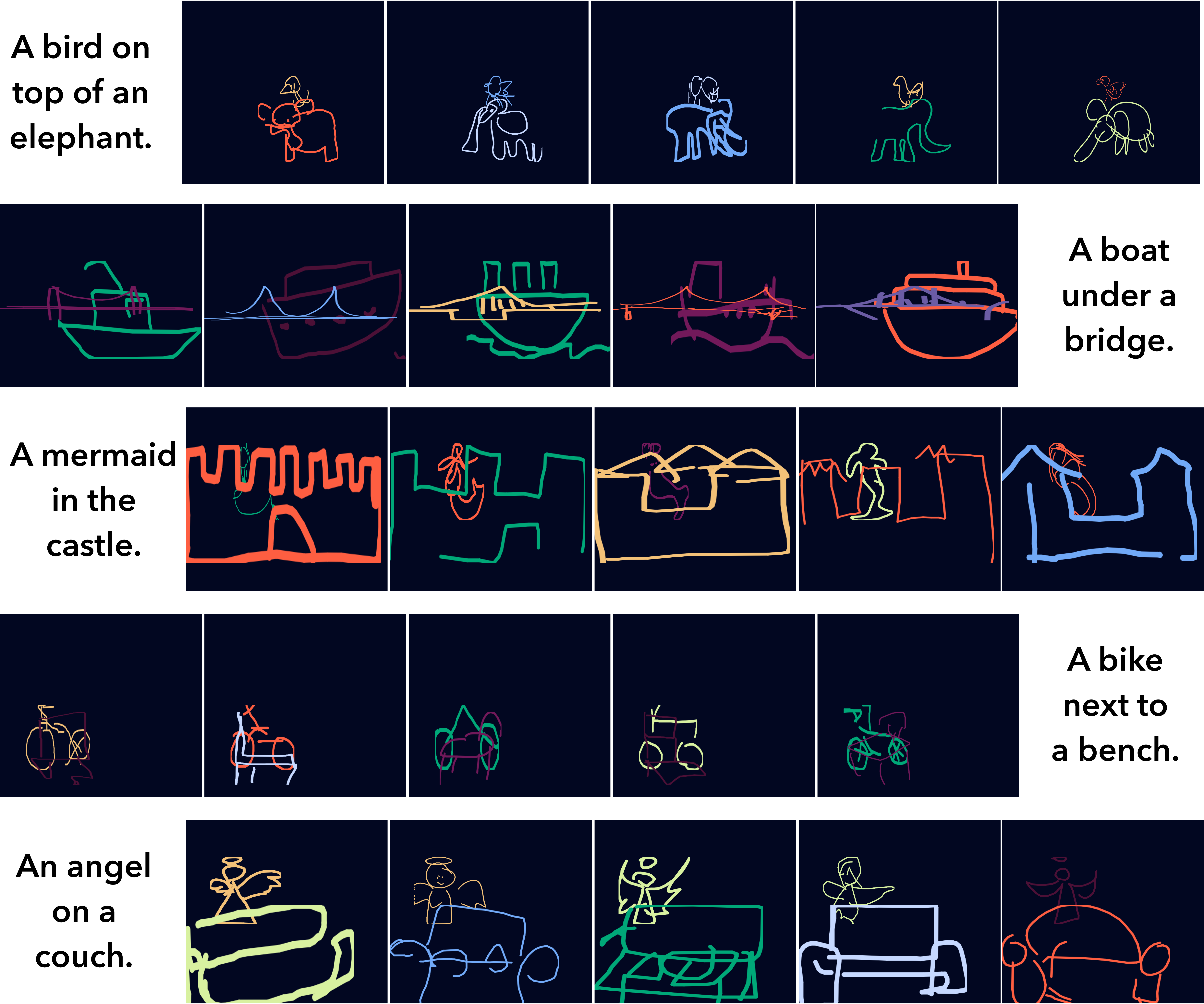}
\end{center}
\caption{Visualization for a simple scene. Apart from the complex scene, our system also supports simple text-guided sketch generation.}
\label{fig:res_simple}
\end{figure}

Though targeting complex scene generation, in Fig. \ref{fig:res_simple} we show that our visualization scheme could also deal with simple text guidance including only positional relations and achieve competing results with previous text guided sketch generation methods\cite{Huang2019}. This function could potentially help with AI-aided sketching with simple text instructions, with which we will further integrate into our interactive storytelling interface in future.

\section{Conclusion}

In this paper, we presented an intelligent visualization interface mainly focusing on fairy tales. Fairy tales, different from casual languages, have unique characters and relationships, which would generally lead direct text-to-image visualization methods to fail. Based on this point, we introduce the doodler graph as an intermediate representation. We fuse the rule-based language model with a learning-based generation model into a unified visualization system to allow for more flexibility during generation. With our fused framework, AI.R Taletorium is able to visualize complex fairy tale scenes with stylized characters and rare relationships and automatically update the visualization along with story generation. 

However, the visualization quality is still constraint by Sketch-RNN generation quality, which could generates unrecognizable characters and making the visualization hard to understand. Re-training the sketcher with Transformer based architecture could help in improving generation quality. In future work, we will re-design the RNN sketcher to improve the doodling quality and further improve the generation efficiency to support real-time mobile inference.

%%
%% The next two lines define the bibliography style to be used, and
%% the bibliography file.
\clearpage
\bibliographystyle{ACM-Reference-Format}
\bibliography{sample-base}

%%
%% If your work has an appendix, this is the place to put it.
\appendix

\end{document}